\documentclass[a4paper,twocolumn,10pt,accepted=2024-06-17]{quantumarticle}
\pdfoutput=1
\usepackage[utf8]{inputenc}
\usepackage[english]{babel}
\usepackage[T1]{fontenc}
\usepackage{amsmath}
\usepackage{hyperref}
\usepackage[numbers]{natbib}

\usepackage{braket}
\usepackage{booktabs}
\usepackage{amssymb}

\begin{document}

\title{Reduction of finite sampling noise in quantum neural networks} 

\author{David A. Kreplin}
\email{david.kreplin@ipa.fraunhofer.de}
\orcid{0000-0002-8129-6864}
\affiliation{Fraunhofer Institute for Manufacturing Engineering and Automation (IPA), Nobelstraße 12, D-70569 Stuttgart, Germany}

\author{Marco Roth}
\email{marco.roth@ipa.fraunhofer.de}
\orcid{0000-0002-1276-5655}
\affiliation{Fraunhofer Institute for Manufacturing Engineering and Automation (IPA), Nobelstraße 12, D-70569 Stuttgart, Germany}

\maketitle

\begin{abstract}
Quantum neural networks (QNNs) use parameterized quantum circuits with data-dependent inputs and generate outputs through the evaluation of expectation values.
Calculating these expectation values necessitates repeated circuit evaluations, thus introducing fundamental finite-sampling noise even on error-free quantum computers.
We reduce this noise by introducing the variance regularization, a technique for reducing the variance of the expectation value during the quantum model training.
This technique requires no additional circuit evaluations if the QNN is properly constructed.
Our empirical findings demonstrate the reduced variance speeds up the training and lowers the output noise as well as decreases the number of necessary evaluations of gradient circuits. 
This regularization method is benchmarked on the regression of multiple functions and the potential energy surface of water.
We show that in our examples, it lowers the variance by an order of magnitude on average and leads to a significantly reduced  noise level of the QNN. 
We finally demonstrate QNN training on a real quantum device and evaluate the impact of error mitigation.
Here, the optimization is feasible only due to the reduced number of necessary shots in the gradient evaluation resulting from the reduced variance. 
\end{abstract}

\section{Introduction}
The methods of quantum machine learning are among the most promising approaches to achieve a quantum advantage in today's era of noisy intermediary quantum (NISQ) computers~\cite{Preskill.2018,Cerezo.2021,Bharti.2022}.
Quantum machine learning (QML) offers significant flexibility in the design of quantum circuits, allowing for the use of short and hardware-efficient circuits.
These circuits may operate more effectively on noisy hardware compared to problem-driven approaches as for example quantum optimization or quantum chemistry. 

In this young field, mainly two approaches are followed to develop machine learning algorithms on NISQ hardware.
The first approach utilizes the exponentially large Hilbert space that is accessible by a quantum computer 
for the so called kernel trick~\cite{Scholkopf.2000}. Here, the input data is embedded into a high dimensional space enabling linear regression or classification within this space~\cite{Theodoridis.2008}.
On a quantum computer, this is achieved by using a quantum feature map~\cite{Schuld.2019b,Lloyd.2020},  a circuit in which data is encoded into a quantum state, for example by using rotation gates~\cite{Havlicek.2019}. Optionally, the quantum feature map may contain additional trainable parameters that enable an adaption of the feature map for the given data~\cite{Hubregtsen.2022, Kubler.2021, Rapp.2024}.
The feature map is also utilized in the second QML approach, in which the output of the QML model are either probabilities of measuring qubits in a certain computational state or expectation values of manually chosen operators~\cite{Mitarai.2018,Bharti.2022}. 
This approach follows the principles of variational quantum algorithms~\cite{Cerezo.2021}, and in this context, the feature map is more commonly referred to as a parameterized quantum circuit (PQC)~\cite{Benedetti.2019}. 
It is also possible to use quantum states directly as an input and manipulate the state by a trainable circuit before measuring it~\cite{Farhi.2018,Cong.2019,Beer.2020,Zhang.2020}.

In the following we denote the second QML approach as Quantum Neural Networks (QNNs) since it is the most established name today~\cite{Cerezo.2022}.
Parameters of QNNs are often optimized by minimizing a loss function similar to the training of a classical artificial neural network (ANN) by gradient descent~\cite{Mitarai.2018,Wierichs.2020,Harrow.2021}.
Differentiation of the expectation value is possible for example by the parameter-shift rule, in which the analytic derivative is obtained through evaluations of the expectation value with shifted parameter values~\cite{Mitarai.2018,Schuld.2019,Wierichs.2022}.

Recent research on QNNs has placed a strong emphasis on understanding their mathematical properties and investigating their advantages and disadvantages.
Like ANNs, QNNs also offer universal function approximation~\cite{Goto.2021, Schuld.2021}, which can already be achieved on the single-qubit level~\cite{PerezSalinas.2021}.
Furthermore, QNNs can reach a high expressivity by the repeated encoding of the input data, also known as data re-uploading~\cite{Vidal.2020,PerezSalinas.2020,Schuld.2021}.
An open question is whether the high expressivity of QNNs is more of a problem than a feature, as it can also easily lead to overfitting of the data. 
On the other hand, research has also demonstrated that QNNs can achieve good generalization~cite{Peters.2023} with only few data points~\cite{Caro.2022}.
Another important topic in the study of QNNs is the phenomenon of barren plateaus, which refers to the observation that the variance of the gradient vanishes exponentially. 
The issue is a well-known challenge~\cite{McClean.2018,Arrasmith.2021} caused by various sources~\cite{Holmes.2022,Marrero.2021,Cerezo.2021b, Uvarov.2021}.
While specially designed PQCs have shown promise in avoiding this issue~\cite{Pesah.2021,Larocca.2023}, another source of barren plateaus arises from noise that can occur during evaluation on real hardware~\cite{Wang.2021,Wang.2024}.
Despite hardware-related noise potentially being reduced in the future, there will always be some level of noise resulting from the finite sampling of the quantum state. 

In this work, we address the challenge of handling finite sampling noise in QNNs within the constraints of the NISQ era, where a massive number of repeated evaluations of a quantum circuit is currently not feasible. 
Today, evaluating a circuit with a larger number of shots comes with long execution times and high financial costs, and therefore, the number of shots is often fixed to a maximum number.\footnote{For example at IBM's currently available hardware, the maximum number of shots is limited to 100$~\!$000 shots. An execution of a single circuit on one of fastest platforms available at this provider with this number of shots takes about 2 minutes. (Experiment: The circuit as displayed in Figure~\ref{fig:pqc} with 20 qubits and 3 layers is evaluated with 100$~\!$000 shots on \texttt{ibm\_cairo})}
As this is a technological issue that is not provider-specific and given the fact that the optimization of a QNN typically requires the evaluation of thousands of circuits, finite sampling noise can be a significant obstacle in training QNNs on real hardware.
Additionally, access to quantum computers is often granted by a pay-per-shot plan~\cite{braket_pricing}, which can result in considerable costs for the full training of a QNN. Therefore, reducing the number of shots is a practical and crucial goal in the current NISQ era of quantum computing.

The standard deviation (STD) associated with the finite sampling noise of the expectation value $E[\hat{H}] = \braket{\Psi|\hat{H}|\Psi}$ of some operator $\hat{H}$ is calculated as:
\begin{equation}
\text{std} (E[\hat{H}]) = \sqrt{\frac{\text{var}(E[\hat{H}]) }{N_\text{shots}}}, \label{eq:sd}
\end{equation}
where $\text{var}(E[\hat{H}])$ represents the variance of the expectation value. 
The formula shows that the finite sampling noise of the expectation value decreases with $O(1/\sqrt{N_\text{shots}})$ as the number of shots $N_\text{shots}$ increases.
However, since the number of shots is practically limited in current NISQ devices, we propose an alternative approach that reduces $\text{std} ( E[\hat{H}])$ by lowering the variance of the expectation value. 
In QNNs, the PQC that generates the wavefunction $\ket{\Psi}$ and the operator $\hat{H}$ are free to choose, and they can be optimized to reduce the variance. We show that a significant reduction of the finite sampling noise can be achieved by adding the variance of the QNN as a regularization term in the loss function. This approach ensures that the training of the QNN not only optimizes the fitting loss but also significantly reduces the variance of the output.

The paper is structured as follows:
In Section~\ref{sec:QNN}, we provide a technical introduction to QNNs and our concept of variance regularization.
Section~\ref{sec:intro_example} presents an introductory example that demonstrates the impact of variance regularization on a selected problem.
Next, in Section~\ref{sec:optimization}, we delve into the details of how variance regularization can be incorporated into the optimization procedure.
We further discuss more deeply the optimization for a regression of the logarithm function and present results for two other functions.
Additionally, we apply the variance regularization in the interpolation of the potential energy surface of water in Section~\ref{sec:water}.
Section~\ref{sec:hardware} focuses on evaluating the performance of QNNs on real quantum computing hardware, and we provide an overview of the optimization procedure conducted on the real hardware.
Finally, we examine the combination of variance regularization and error mitigation techniques, specifically zero-noise extrapolation.

\section{Variance Regularization \label{sec:QNN}}

The QNN employed in this works utilizes a PQC, denoted as $\hat{U}$, to encode the classical input data $x \in \mathbb{R}$ into a quantum state $\ket{\Psi}$ and manipulate the quantum state using additional parameters $\varphi$:
\begin{equation}
\ket{\Psi(x,\varphi)} = \hat{U}(x,\varphi)\ket{0}.
\end{equation}
Practically, the PQC $\hat{U}(x,\varphi)$ is primarily constructed from one- and two-qubit gates, and often rotational gates are utilized for encoding data and manipulating the quantum state. 
Employing repeated input encoding techniques enhances the expressibility of the circuit~\cite{PerezSalinas.2020}. An example of a PQC used in this study can be found in Figure~\ref{fig:pqc}.

The output of the QNN, denoted as $f(x)$, is obtained by evaluating the expectation value of a cost operator $\hat{C}(\phi)$ as follows:
\begin{equation}
f_{\varphi,\phi}(x) = \braket{\Psi(x,\varphi)|\hat{C}(\phi)|\Psi(x,\varphi)}. \label{eq:QNN}
\end{equation}
In scenarios where multiple outputs are desired from the QNN, an individual cost operators $\hat{C_j}(\phi_j)$ can be used for each output. In such cases, the QNN output $f$ is considered as a vector containing the expectation values of each respective cost operator.
When there is no need to distinguish between the parameters $\varphi$ of the PQC and the parameters $\phi$ of the cost operators, they can be combined into a single parameter vector $\theta$. For convenience, the explicit dependency on the parameters is often neglected in the subsequent discussions.

The parameters $\theta$ of the QNN are determined through the minimization of a loss function $L_\text{fit}(\theta)$.
The choice of the loss function depends on the specific task at hand and plays a crucial role in the model's performance.
It is common to adapt and utilize classical machine learning loss functions within the context of QNNs.
In this work, we mainly focus on regression for which a possible loss function is given by the squared error (named in the following fitting loss):
\begin{equation}
L_\text{fit}(\theta) = \sum_i ||f_\theta(x_i)-y_i||^2. \label{eq:fit_loss}
\end{equation}
Here, we consider a set of input data $\{x_i \in \mathbb{R}^d\}$ and their corresponding labels $\{y_i \in \mathbb{R}\}$ within a supervised learning scenario. 
In case of a classification task the cross-entropy loss function is often employed~\cite{Skolik.2021}.
Moreover, alternative loss functions empower the QNN to tackle other problems such as differential equations~\cite{Kyriienko.2021}.

The optimization of the parameters $\theta$ involves minimizing the loss function which can be challenging because the loss function is typically non-convex.
Therefore, gradient-based optimization methods are commonly employed to address this task~\cite{Harrow.2021}.
To obtain derivatives with respect to the parameters $\varphi$ of the PQC, one approach is the parameter-shift rule~\cite{Mitarai.2018, Schuld.2019, Wierichs.2022}.
In this method, the parameter is shifted in each gate by a specific value and the analytic derivative is obtained by the difference of the resulting expectation values. 
Consequently it is necessary to evaluate two circuits for each gate containing the parameter.
However, when dealing with PQCs that consist of numerous parameterized gates, the evaluation of the resulting amount of circuits can become a computational bottleneck, particularly when utilizing state-of-the-art quantum hardware.
While alternative approaches, such as the linear combination of unitaries do exist~\cite{Somma.2002, Guerreschi.2017}, it is important to consider that these approaches may be more vulnerable to hardware noise when compared to the parameter-shift rule.

The differentiation with respect to the cost operator parameters $\phi$ is straightforward:
\begin{equation}
\partial_{\phi_m} f_\theta(x) = \braket{\Psi(x,\varphi)| \big(\partial_{\phi_m} \hat{C}(\phi)\big)|\Psi(x,\varphi)}.
\end{equation} 
The derivative $\partial_{\phi_m} \hat{C}(\phi)$ of the cost operator can be often evaluated from the same measurement as the expectation value of the cost operator.
In our routines, we reuse the results from the circuit evaluation of the cost operators to calculate these derivatives. 
 
Once the loss function is minimized to a satisfactory low value, the QNN is considered to be trained and the model can be used to obtain new predictions by inputting new data into Eq.~\eqref{eq:QNN}.
However, it is important to keep in mind that the output of the model is determined by the stochastic process of measuring the expectation value, and therefore, in contrast to classical ANNs, the output contains noise.
The standard deviation of the expectation value, i.e. the model output, is obtained by (cf. Eq.~\eqref{eq:sd}):
\begin{equation}
\text{std}(f) = \sqrt{\frac{\sigma^2_{f}}{N_\text{shots}}}. \label{eq:sd2}
\end{equation}
The variance $\sigma^2_{f}$ of the QNN can be computed by 
\begin{equation}
\sigma^2_{f} = \braket{\Psi|\hat{C}^2|\Psi} - \braket{\Psi|\hat{C}|\Psi}^2. \label{eq:var}
\end{equation}
In case of multiple outputs, the standard deviation is obtained for each output by evaluating the variance for each cost operator $\hat{C}_j$ separately. 

In the following we utilize the variance as a regularization during the optimization of the QNN in order to reduce the noise of the model output. 
This is achieved by adding the following term to the loss function that minimizes the variance of the function at a set of points $\{\tilde{x}_k\}$:
\begin{equation}
L_\text{var}(\theta) =  \sum_k || \sigma^2_f(\tilde{x}_k) ||, 
\end{equation}
where $||.||$ is a suitable vector norm. 
The total loss that is minimized during the training procedure is the sum
\begin{equation}
L = L_\text{fit} + \alpha L_\text{var}. \label{eq:varparam}
\end{equation}
The hyper-parameter $\alpha>0$ is used to adjust the balance between the fitting error and the variance reduction of the model. 
In our experience, a value between $10^{-2}$ and $10^{-4}$ yields the satisfying results.
The choice of $\{\tilde{x}_k\}$ is in principle arbitrary and should evenly capture the domain space of the input data.
We assume that the training data is a good representation of the model domain, and therefore, we will set $\{\tilde{x}_k\}=\{x_i\}$ in the following. 
This approach offers the advantage of reusing the circuits and function evaluations of $f(x)$ from the fitting loss $L_\text{fit}$.

The computationally complexity of $\braket{\Psi|\hat{C}^2|\Psi}$ strongly depends on the choice of the $\hat{C}$.
If the operator is idempotent, i.e. $\hat{C}=\hat{C}^2$, the evaluation of this term comes with no additional costs. For example, this is the case for the probability $p_n(0)$ (or $p_n(1)$) of a single qubit $n$, since this is equivalent to a cost operator of $\ket{0}\!\!\bra{0}_n$ (or $\ket{1}\!\!\bra{1}_n$) in which $n$ labels the qubit.  
When the operator is diagonal in the computational basis, it becomes feasible to evaluate the expectation value of the squared operator using the same circuit measurements as those employed for function evaluation. This scenario occurs for example when the operator solely consists of Pauli operators $\hat{I}$ and $\hat{Z}$, as observed in Ising Hamiltonian with the following form:
\begin{equation}
\hat{C}  = \phi_1 \hat{I} + \phi_2 \sum_p \hat{Z}_p + \phi_3 \sum_ {p>q} \hat{Z}_p \hat{Z}_q\,. \label{eq:cost_op_intro}
\end{equation}
This is also true for the evaluation of the gradient of the variance:
\begin{equation}
\nabla_\theta \sigma^2_f = \nabla_\theta \braket{\Psi|\hat{C}^2|\Psi} + 2 \braket{\Psi|\hat{C}|\Psi} \nabla_\theta \braket{\Psi|\hat{C}|\Psi}\,.
\end{equation} 
The second term is computed from the intermediates that are also used in the gradient computation of the loss function $L_\text{fit}$.
The derivatives of the first term are obtained using the same circuits and measurements as those used for calculating the gradient of the function.
Extra work is purely introduced in the classical post-processing to evaluate the additional expectation values from the measurements. 

For general observables, the operators $\hat{C}$ and $\hat{C}^2$ can be simultaneously measured in the eigenbasis of $\hat{C}$.
Therefore, in principle, the quantum computation overhead is only increased by the additional unitary transformation to the eigenbasis.
If this eigenbasis is unknown or impractical to compute, evaluating the squared cost operator requires a separate calculation of the expectation value $\braket{\Psi|\hat{C}^2|\Psi}$.
This may make the application of variance regularization impractical in variational algorithms involving many non-commuting terms in the observables, as is often encountered in the variational quantum eigensolver (VQE).
To discuss this in more detail, we consider the following transverse Ising Hamiltonian:
\begin{equation}
\hat{C} = -J \sum_{p>q} \hat{Z}_p\hat{Z}_q - h \sum_p \hat{X}_p.
\end{equation}
The observables involving $\hat{Z}_p\hat{Z}_q$ and $\hat{X}_p$ can be organized into distinct groups of commuting observables. Consequently, evaluating the expectation value of $\hat{C}$ requires measurements from two separate circuits.
However, in the case of the squared observable $\hat{C}^2$, the count of distinct groups with commuting observables increases to $2N_q +2$ with $N_q$ qubits. This expansion includes a group solely with pure $\hat{Z}$ observables, another group with pure $\hat{X}$ observables, and $2$ groups per qubit, each involving either $\hat{X}$ or $\hat{Y}$ acting on the qubit while $\hat{I}$ and $\hat{Z}$ operate on all other qubits.
Hence, $2N_q +2$ circuits must be measured, introducing a significant overhead for calculating the variance of the given transverse Ising Hamiltonian.

In QNNs, however, the situation is considerably different since the choice of the observable is usually flexible.
Commonly used observables in the literature utilize expectation operators of Pauli-Z observables or probabilities of being in a certain computational state~\cite{Cerezo.2021,PerezSalinas.2021,Pesah.2021,Skolik.2021,Goto.2021,Caro.2022}.
The remainder of the manuscript thus explores scenarios where C is diagonal in the computational basis.

\section{Introductory Example \label{sec:intro_example}}

In this section, we present an illustrative example that demonstrates how variance regularization significantly enhances the robustness of a QNN against finite-sampling noise within a shot-based simulation.
This example also highlights that a perfect fit, derived from a noise-free simulation, could inadvertently introduce a substantial variance of the expectation value.
Consequently, this would necessitate a significantly large number of shots for an accurate evaluation.
We want to point out that  this effect should be taken into account when interpreting results from QNNs that are entirely sourced from noise-free simulators.

\begin{figure}[t]
\includegraphics[width=1\linewidth]{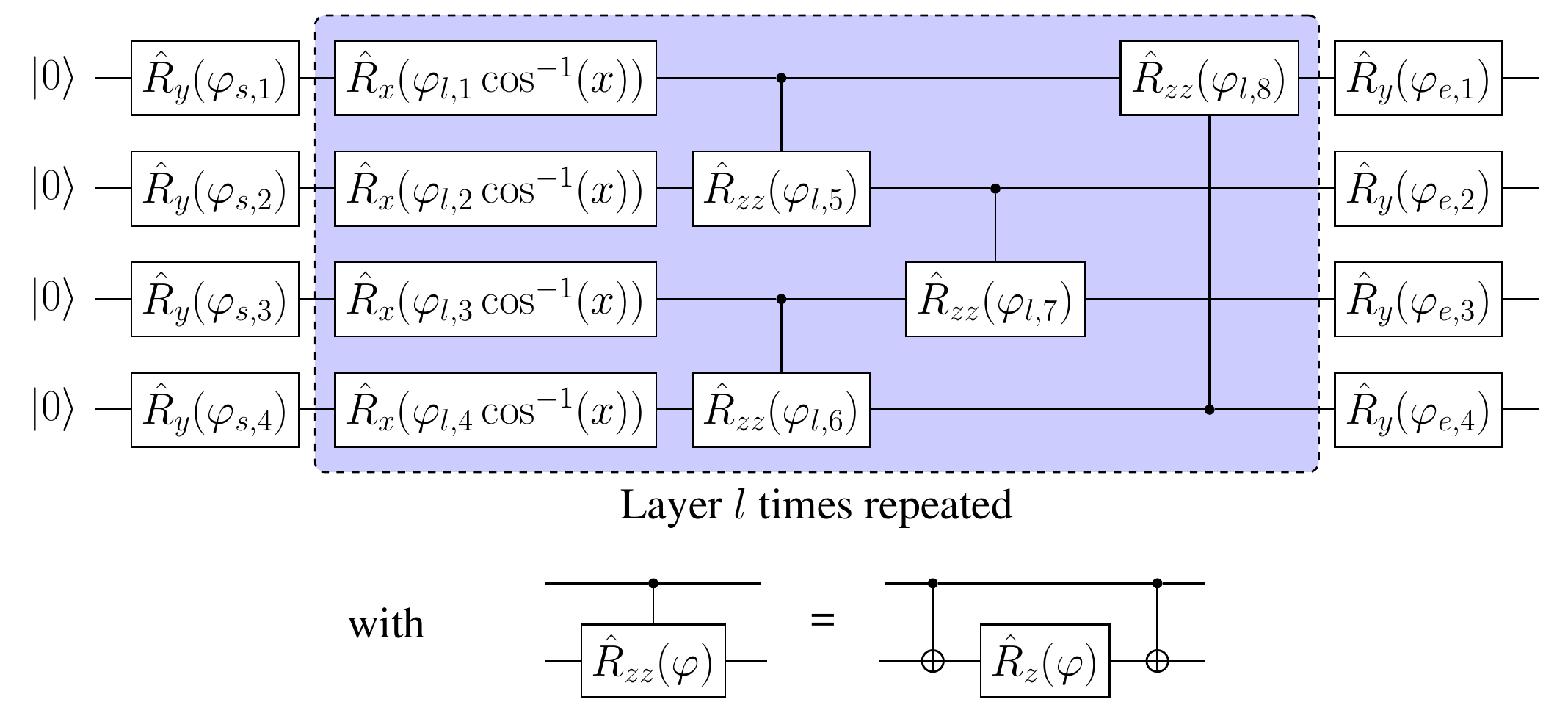}
\caption{Parameterized quantum circuit of the QNN used in all examples in this work.
The first layer of $\hat{R}_y$ gates manipulates the initial state. The blueish highlighted layer includes the Chebychev input encoding in the $\hat{R}_x$ gates as well as the parameterized control manipulation of the quantum state. 
The layer is repeated $l$ times for a repeated input encoding. The last layer of $\hat{R}_y$ gates serves as a change of the basis that is used for measurement. 
For a hardware efficient approach, the rightmost controlled gate in the blueish layer is removed to avoid swapping. \label{fig:pqc}
}
\end{figure}

Figure~\ref{fig:pqc} displays the PQC that is utilized throughout this work.
We use the Chebyshev input encoding in which the input data is first encoded by a $\cos^{-1}(x)$ function~\cite{Mitarai.2018}.
The use of the inverse cosine implies a rescaling of the input data to $[-1,1]$.
Using this non-linear input encoding as angles in rotation gates around the x-axis yields the following identity for integer $n$~\cite{Kyriienko.2021}:
\begin{equation}
\hat{R}_x(n \cos^{-1}\!(x))\! = \!T_n(x) \hat{I} - i\sqrt{1-x^2} U_{n-1}(x)\hat{X}\,,
\end{equation}
in which $T_n(x)$ and $U_n(x)$ are the first and second kind Chebyshev polynomials of degree $n$~\cite{Abramowitz.1988}. $\hat{X}$ denotes the Pauli-X operator.
In contrast to previous works, we introduce a parameter $\varphi$ instead of a fixed value for $n$ which enables a flexible optimization of the degree of the Chebyshev polynomials during the training.
An example of curves generated using Chebyshev polynomials with a non-integer value of $n$ is presented in the Appendix~\ref{sub:cheb}. These curves demonstrate a smooth transition between the individual Chebyshev polynomials.

The initial values for the parameters in the encoding are chosen to be evenly distributed between $\varphi_{l,1}=0.01$ and $\varphi_{l,{N_q}}=\beta$ following the idea of the Chebyshev tower approach~\cite{Kyriienko.2021}.
Here and in the subsequent sections, $N_q$ denotes the number of qubits.
The initial value 0.01 is chosen to avoid a redundancy in the parameter that occurs for $\varphi=0.0$, and $\beta$ is considered as a hyper-parameter of the model.
The quantum state from the input encoding is manipulated by a hardware efficient approach of $\hat{R}_{zz}(\varphi) = \exp(-i \tfrac{\varphi}{2} \hat{Z} \otimes \hat{Z})$ two-qubit interaction gates that are arranged in a nearest neighbor entangling set-up. 
The $\hat{R}_y$ gates at the beginning and the end of the feature map enable a basis change of the initial state and the measuring basis. Both are also optimized during the training. 

\begin{figure}[t]
\centering
\vspace{2mm}
\includegraphics[width=\linewidth]{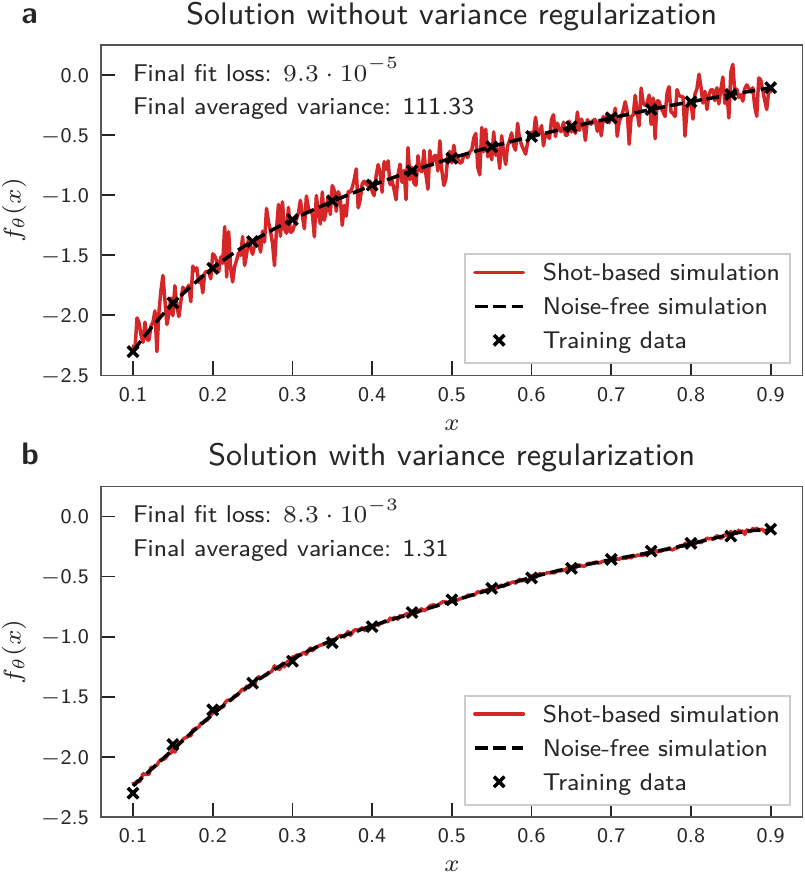}
\caption{
The output of the QNN is evaluated for two cases: trained with (\textbf{a}) and without (\textbf{b}) variance regularization. The training is conducted using a noise-free simulator, and the output is computed with and without shots. For the shot-based simulation, 10$~\!$000 shots are utilized.
  \label{fig:intro_example} }

\end{figure}

In this introductory example, we choose the Ising Hamiltonian introduced in Eq.~\eqref{eq:cost_op_intro} as the cost operator.
The parameters $\phi_1$ to $\phi_3$ of the Ising Hamiltonian are also optimized in the optimization.
The first parameter $\phi_1$ introduces a constant offset of the output of the QNN. 
The specific selection of this operator exemplifies a scenario where a large variance is observed after the training process.
The PQC of the QNN is introduced in Figure~\ref{fig:pqc}, and $N_q=4$ qubits and $l=2$ layers are chosen. 
The optimization of the loss function, i.e. the training, is conducted using the noise-free simulator of Qiskit~\cite{Treinish.2023}, and it is performed with and without the variance regularization. 

Figure~\ref{fig:intro_example} displays the inference of the trained QNN with and without finite sampling noise. 
The top plot illustrates the QNN obtained without variance regularization, while the plot at the bottom  demonstrates the outcomes with variance regularization.
Note that the noise free output of the QNN, represented by the dotted black line, accurately reproduces the logarithm function in both plots.
The results of the shot-based simulator, here the QASM simulator in Qiskit~\cite{Treinish.2023}, are computed utilizing 10$~\!$000 shots.
By incorporating the variance regularization, the averaged variance of the trained QNN is reduced significantly by a factor of 85.
This reduction in variance impacts the model's precision, and it leads to an increase in the squared error from $9.3\cdot 10^{-5}$ to $8.3\cdot 10^{-3}$ in the noise-free optimization.
However, there is no substantial visual difference observed in the noise-free outcome for these low fitting losses.

This picture changes when considering the results of the shot-based simulations.
Here, the difference between the evaluation of the QNN with the parameters obtained with and without variance regularization during the training is substantial. While the former results in a drastically noisy inference, the latter is difficult to visually distinguish from the noise-free result.
Effectively, the 85-fold reduction in variance implies that the same level of shot noise can be achieved with 85 times fewer shots. Alternatively, if the number of shots is kept constant (as depicted in Figure~\ref{fig:intro_example}), the standard deviation (cf. Eq.~\eqref{eq:sd2}) is reduced by a factor of 9.2. 
The example without variance regularization further illustrates that the QNN obtained through training with a noise-free simulator may not be practical for evaluations on a shot-based backend. Despite its high accuracy, the solution without variance regularization would require approximately 850$~\!$000 shots to achieve a similar result as the QNN trained with variance regularization.

The cost operator presented in Eq.~\eqref{eq:cost_op_intro} exhibits limited flexibility due to its inclusion of only three free parameters. Additionally, the two-body interaction term is not well-suited for calculating the variance, as classical post-processing scales with $O(N_q^4)$.
In the subsequent sections, we will utilize a different operator for our experiments, given by:
\begin{equation}
\hat{C}(\phi) = \phi_0 \hat{I} + \sum_p^{N_q} \phi_p \hat{Z}_p. \label{eq:sumZ}
\end{equation}
Our experiments have shown this operator to be more versatile and less susceptible to shot noise in general. Therefore, all the results presented in the following sections are obtained using this cost operator. 

\section{Optimization with variance regularization \label{sec:optimization}}

\begin{figure}[t]
\vspace{2mm}
\centering
\includegraphics[width=\linewidth]{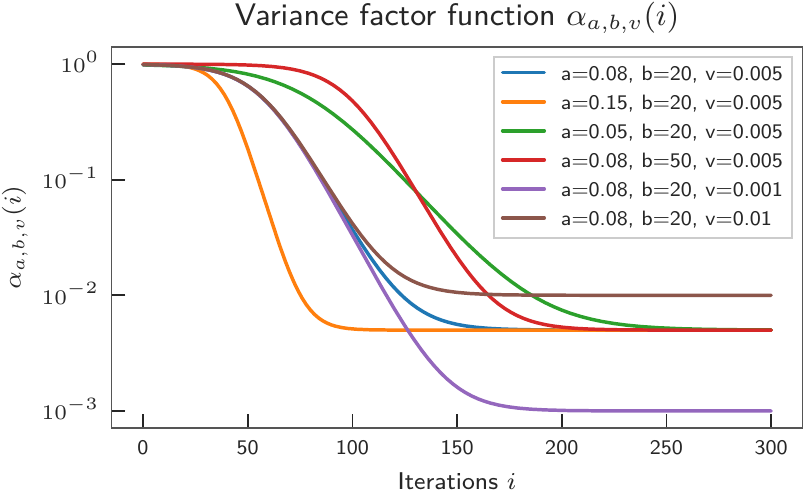}
\caption{Different regularization parameter functions $\alpha_{a,b,v}(i)$ for various combinations of $a$, $b$ and $v$. The blue curve with $a=0.08, b=20, v=0.005$ is used for the experiments throughout this paper.  \label{fig:varfac}}
\end{figure}

In this section, we show numerical evidence for the benefits of using the variance regularization approach introduced in Section~\ref{sec:QNN}.
All following calculations have been executed with the QML python package sQUlearn~\cite{Kreplin.2023}, in which the variance regularization has been integrated.
In the following optimization we compare two approaches for choosing the parameter $\alpha$ in the variance regularization (cf. Eq.~\eqref{eq:varparam}). 
In the first approach, the parameter remains constant throughout the optimization process.
The second approach involves adjusting the parameter dynamically during optimization.
The objective is to initially force the QNN to significantly reduce the variance and then gradually transition towards a regime where the squared error (cf. Eq.~\eqref{eq:fit_loss}) dominates the total loss.
This approach results in a significantly lower final variance value.
An additional benefit of reducing the variance at the beginning is the ability to use a lower number of shots in the initial iterations.
The parameter $\alpha$ is determined using a modified sigmoid function, given by:
\begin{equation}
\alpha_{a,b,v}(i) = (1-v)\frac{b\exp(a(b-i))}{b\exp(a(b-i))+1}+v. \label{eq:varfunc}
\end{equation}
Here, the parameter $v$ represents the strength of the variance regularization at the end of the training, and $b$ defines the width of the plateau in the initial phase when the optimization primarily focuses on reducing the variance of the function. The parameter $a$ determines the slope of the decay in the regularization parameter.
A plot illustrating the regularization parameter $\alpha_{a,b,v}$ for various parameter values is presented in Figure~\ref{fig:varfac}.
As detailed in Appendix~\ref{sec:parameters}, the fully converged solutions primarily depend on the parameter $v$, which balances variance and fitting loss. Higher values of $v$ reduce the variance more but also increase the fitting loss. In this work, we consistently used a value of $v=0.005$, though different applications might require adjustments. The parameters $a$ and $b$ can be used to adjust the variance reduction at the beginning of training. Parameter $b$ should ensure a long enough plateau to reduce variance when the fitting loss is initially high. Parameter $a$ controls the rate of variance increase after reaching its minimum, however, results are not very sensitive to the precise value of $a$. 
More details on parameter selection and results for different parameters can be found in Appendix~\ref{sec:parameters}.

In addition, the number of shots for evaluating the gradient is adjusted during the optimization.
It is readily apparent that a higher number of shots at the beginning of the optimization is not needed, since a precise evaluation is not beneficial in case of a large fitting error. 
In this section, we introduce a procedure that automatically adjusts the number of shots during the evaluation of the gradient circuits.
It is based on the relative standard deviation (RSD) of the fitting loss $L_\text{fit}$~\eqref{eq:fit_loss}. The number of shots are adjusted such that the RSD is lower than a predefined boundary. 
By using the approximation $\text{var}(f(X)) \approx (f'(E[X]))^2 \text{var}(X)$~\cite{Benaroya.2005} one obtains for the variance of the fitting loss $L_\text{fit}$:
\begin{equation}
\text{var}(L_\text{fit}) = 4 \sum_i (f_\theta(x_i)-y_i)^2 \frac{\sigma^2_f(x_i)}{N_\text{shots}}. \label{eq:var_Lfit}
\end{equation}
Empirically, we have observed that this approach also provides a good approximation of the variance of the total loss function $L$, with the variance of $\text{var}(L_\text{var})$ being notably lower.
During the initial stages of optimization, the fitting error dominates the variance of the total loss.
Furthermore, as the optimization progresses and the regularization parameter $\alpha$ becomes smaller, the contribution of the variance loss to the total variance diminishes to a negligible extent.
Using Eq.~\eqref{eq:var_Lfit}, the relative standard deviation of the fitting error can be expressed as follows:
\begin{equation}
\text{rsd} (L_\text{fit}) = \frac{\sqrt{4 \sum_i (f_\theta(x_i)-y_i)^2 \frac{\sigma^2_f(x_i)}{N_\text{shots}}}}{\sum_i (f_\theta(x_i)-y_i)^2} < \beta.
\end{equation}
The number of shots for evaluating the circuits of the gradient computation is obtained by setting an upper limit of the RSD by a predifined hyper-parameter $\beta$. Additionally, a minimum (100) and maximum number of shots is set.
In our experience, a value of $\beta=0.1$ is enough for a gradient evaluation that does not negatively impact the optimization compared to an optimization with the given maximum number of shots.
\begin{figure*}[t]
\centering
\includegraphics[width=0.98\linewidth]{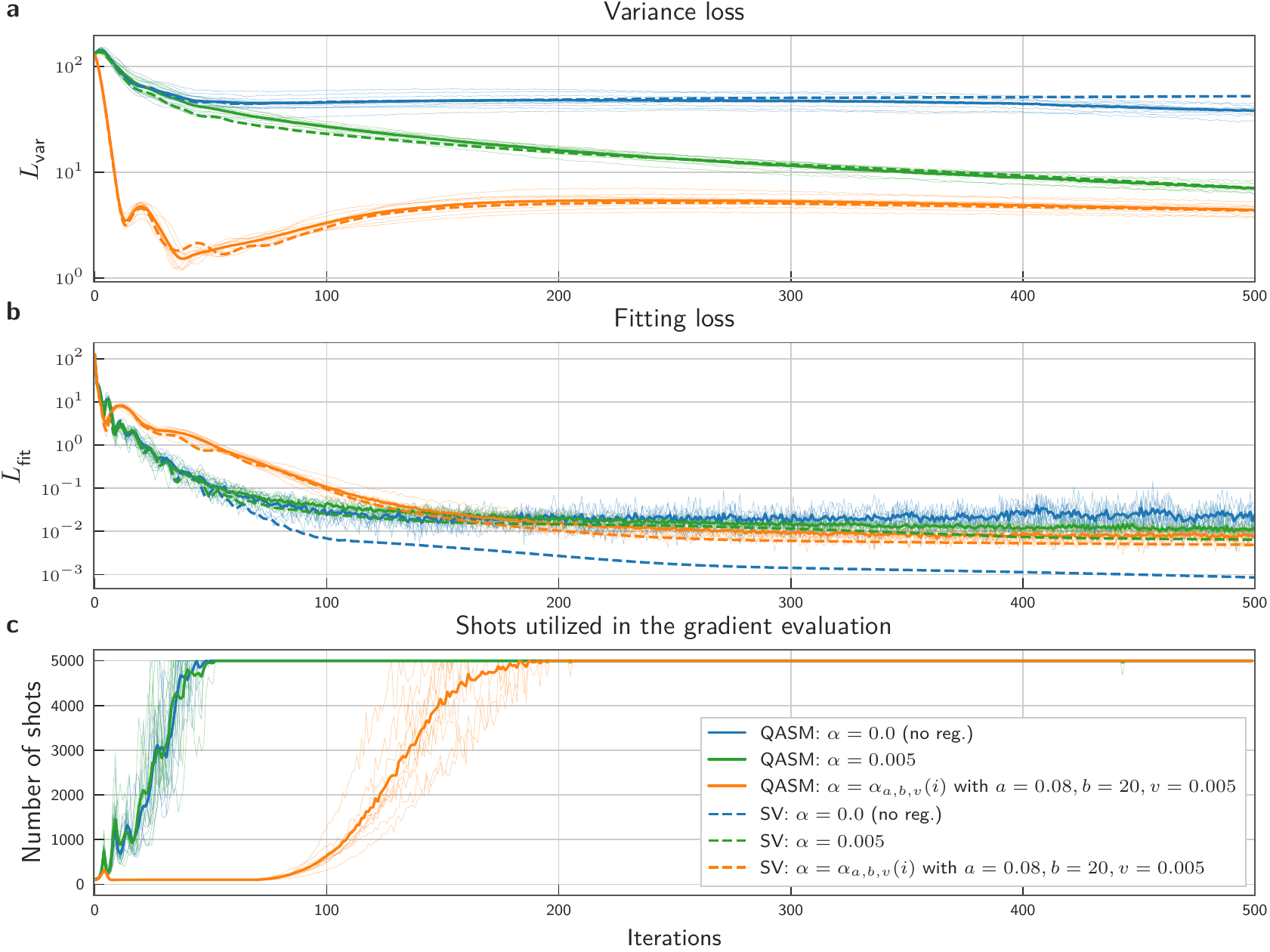}
\caption{Graphs representing the ADAM optimization of a QNN using the shot-based QASM simulator.
The upper panel (\textbf{a}) showcases the variance loss (excluding the prefactor $\alpha$), while the middle panel (\textbf{b}) displays the fitting loss throughout the optimization process.
The bottom panel (\textbf{c}) illustrates the number of shots utilized in the gradient evaluation.
All results are averaged values from 10 runs, and the individual results are depicted as thin lines. Results from the noise-free statevector simulator (SV) are represented by dashed lines. 
 \label{fig:opt_plot}}
\end{figure*}

The criterion for determining the number of shots was designed in such a way that no additional evaluations of quantities are required.
The values of $f_\theta(x_i)$ and $\sigma^2_f(x_i)$ are already included in the loss function and are evaluated with the maximum number of shots.
Subsequently, the number of shots in the evaluation of the circuits needed for the gradient computation is adjusted.
The variance of the loss function, theoretically, does not provide any information about the variance of its gradient.
We are not aware of any boundaries that can be derived for the gradient of the loss function without introducing additional circuit evaluations and expectation value calculations.
As a result, there is no theoretical guarantee for determining the correct number of shots to ensure a predefined noise level for the gradient.
However, empirically, we have found that this approach yields satisfactory results, and it achieves a similar optimization performance compared to working with the maximum number of shots.

\begin{figure*}[t]
\centering
\includegraphics[width=\linewidth]{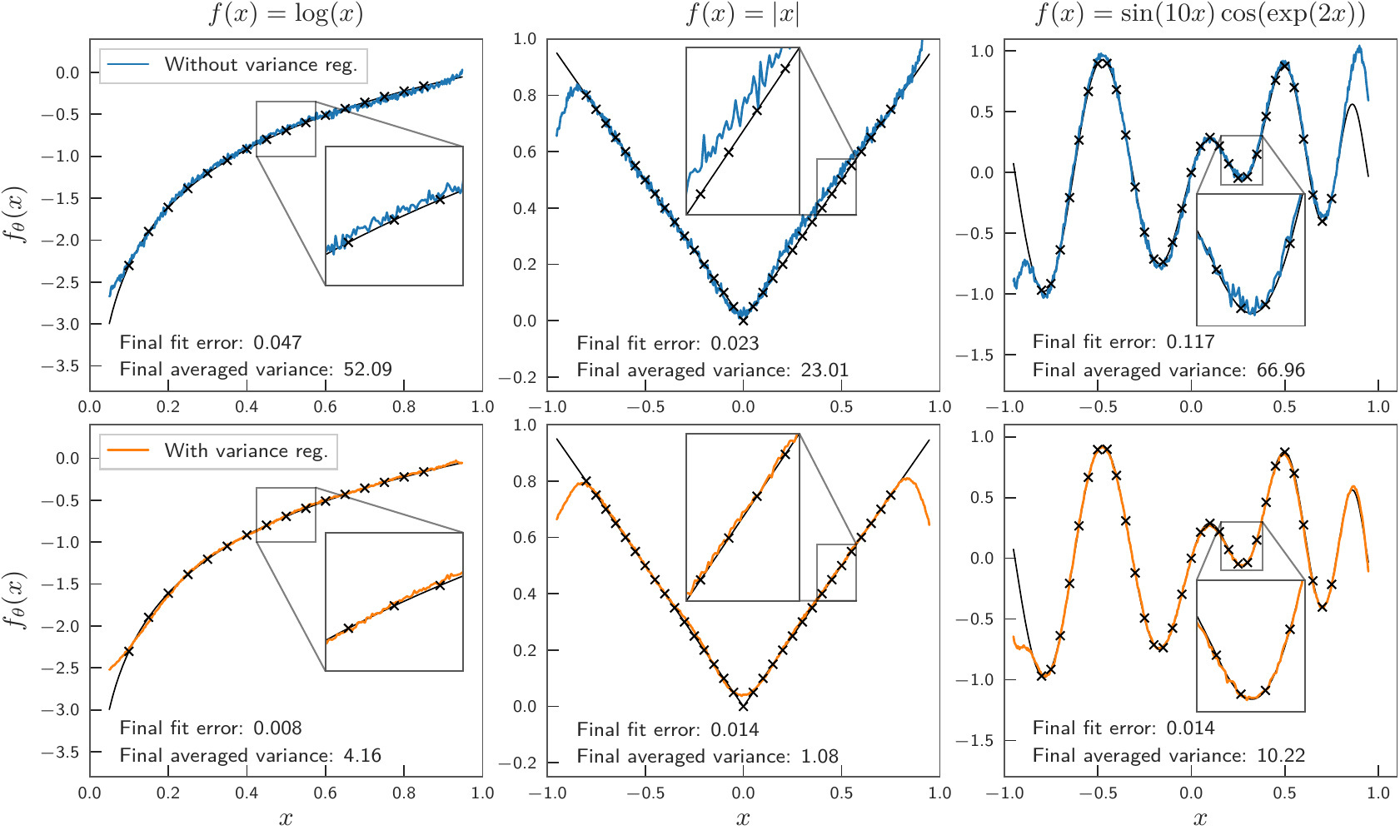}
\vspace{-0.7cm}
\caption{Regression of various functions without (blue) and with (orange) variance regularization. The black lines show the reference function; the data points marked with an x are used for training the QNN. 
The training and the final inference are obtained from a shot-based QASM simulation using 5000 shots.
The inference for the logarithm function is performed on a test set with an equidistant spacing of 0.002 while the spacing for the other functions is 0.004. 
  \label{fig:3results} }
\end{figure*}

Figure~\ref{fig:opt_plot} illustrates the optimization process for fitting the logarithmic function.
Displayed from top to bottom are the variance loss, the fitting loss, and number of shots used in the gradient evaluation.
The left panel of Figure~\ref{fig:3results} presents the training data used in the optimization, and the final fit.
The figure showcases three different scenarios of variance regularization: (i) blue represents no variance regularization, (ii) green corresponds to a constant regularization parameter $\alpha=0.005$, and (iii) orange represents the regularization parameter $\alpha_{a,b,v}$ that changes during the iteration following the blue curve in Figure~\ref{fig:varfac}.
The QNN employed is based on the PQC depicted in Figure~\ref{fig:pqc}.
It utilizes $N_q=10$ qubits, $l=3$ layers, and the cost operator introduced in Eq.~\eqref{eq:sumZ}. 
The optimization is performed by ADAM~\cite{Kingma.2017} with a learning rate of 0.1 utilizing IBM's shot-based QASM simulator with a maximum of 5000 shots.
The dashed lines in the plots represent statevector simulations without any noise. The shot-based optimization is repeated 10 times with the same initial parameters, and the individual results are displayed using thin lines.
The solid lines represent the averaged result over the 10 runs.

We first discuss the results without any variance regularization (blue curves). Without any finite measurement errors, the fit error of the noiseless optimization is the lowest of all optimization strategies. 
However, switching to the shot-based simulator yields a considerably different picture.
The fit error reaches a considerably higher plateau compared to the statevector simulation, primarily due to the finite sampling noise, and the final value is the highest compared to all other methods. 
Nonetheless, the variance slightly decreases as the optimization progresses. We conclude that the ADAM optimization indirectly addresses the variance in an attempt to further minimize the fitting error.
However, this process occurs at a slow and inefficient pace.

Adding the variance regularization with a constant parameter $\alpha$ (green curves) strongly reduced the variance of the QNN, and the final variance is significantly lower.
The fit error in the noise-free simulation is higher compared to the case without variance regularization, as the optimization is not solely focused on minimizing the fit error.
However, this effect reverses when transitioning to the shot-based simulation, in which the reduced variance also improves the convergence to a lower fit error.

Moving on to the iteration-based variance parameter, we observe a similar trentow in the fit error. However, the variance is significantly decreased compared to the constant parameter $\alpha$.
The variance plot clearly demonstrates that this strategy primarily targets the variance reduction first, while prolonging the optimization of the fit.
On the other hand, the lower variance results in fewer shots required for the time-consuming gradient evaluation, leading to a faster overall computation time until reaching the plateau after 300 iterations. The final variance is reduced from 52.09 (without variance regularization) to 4.16.
In other words, the final QNN obtained with variance regularization requires over ten times fewer shots to achieve a similar level of finite sampling noise.

Figure~\ref{fig:3results} presents the final results of regression on three different functions: the logarithm, the absolute value function, and an oscillating function.
The figures also display the averaged values of the final fit error and variance over the last 10 optimization iterations.
The optimization process follows the methodology described above. 
For the absolute value function, the fitting error of the individual training data points is weighted, with higher weights assigned to the points in the center. 
These weights are determined by the function $w(x)=2\exp(-x^2)$.

In all three examples, there is a noticeable reduction in both finite sampling noise and fit error.
Particularly, the example of the absolute value function demonstrates a remarkable reduction in variance, exceeding a factor of 20.
The reduction in fit error is evident in the plot for the oscillating function.
The operator utilized in these examples (cf. Eq.~\eqref{eq:sumZ}) has the capability to generate results without the same level of significant noise observed in the introductory example.
However, despite this inherent capability, the application of variance regularization greatly enhances the results for all three examples, without imposing a notable computational overhead.

\section{Application: Potential energy surface \label{sec:water}}

\begin{figure*}
\vspace{2mm}
\begin{minipage}[t]{0.48\linewidth}
\centering
\includegraphics[width=\linewidth]{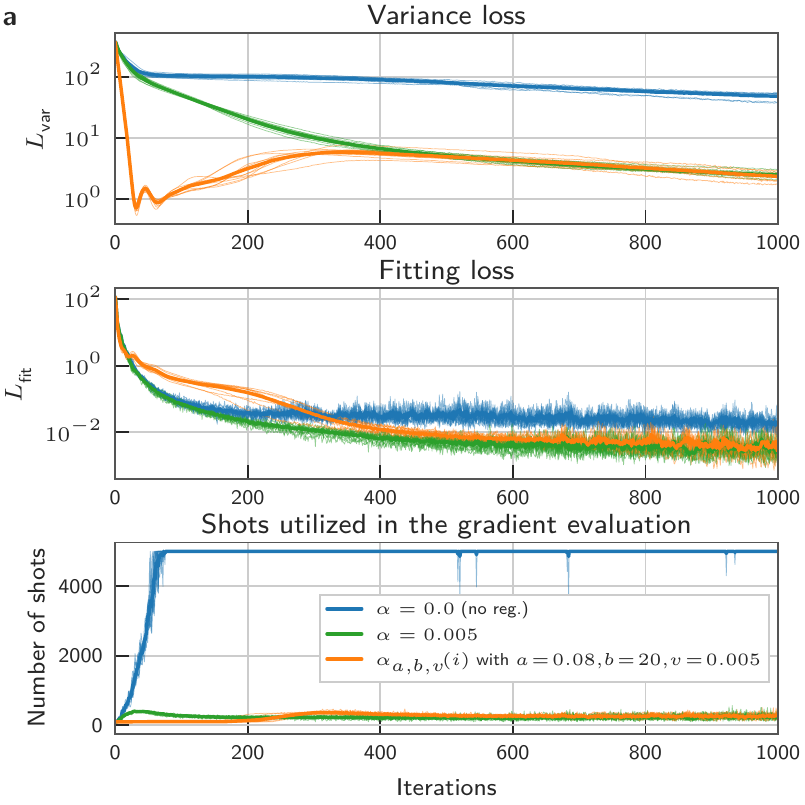}
\end{minipage}
\hfill
\begin{minipage}[t]{0.48\linewidth}
\centering
\includegraphics[width=\linewidth]{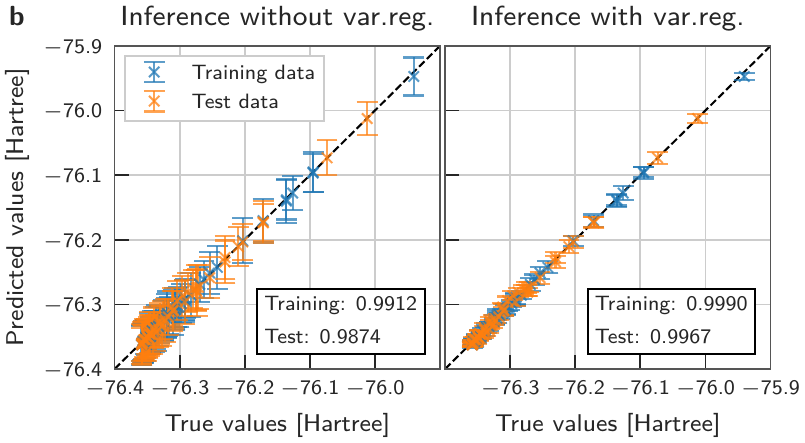}
\end{minipage}
\vspace{-0.1cm}
\caption{\textbf{a}: Results from the optimization of the water PES using a shot-based QASM simulator.
\textbf{b}: Inference of the training and test data with the best trained model. The error bars represent the 95\% confidence interval of the rescaled QNN output. These confidence intervals are obtained by performing 100 evaluations for each data point using 5000 shots. The X-marks indicate the averaged output over the 100 calculations. The $R^2$ scores are shown for training and test data. 
\label{fig:h2o_results}
}
\end{figure*}

In this section, we discuss an application addressing the interpolation of potential energy surfaces (PES) for molecules. PES are derived by solving the electronic Schr\"odinger equation for molecules, considering the spatial coordinates of their constituent atoms. They offer valuable insights into stable molecular configurations, reaction pathways, and reaction kinetics. Computing the energy for a given molecular geometry is computationally demanding with unfavorable scaling, making it crucial to obtain a good representation of the PES from a limited number of data points. Given that the data stems from inherent quantum simulations, this presents an interesting application for quantum machine learning~\cite{Kiss.2022}.

In the following, we demonstrate an interpolation of a PES for the water molecule based on dataset~\cite{water_dataset} provided in Ref.~\cite{Schmitz.2019}. The coordinates of water molecules are transformed into the three degrees of freedom of the molecule: the two distinguished bond distances between the Oxygen and one of the Hydrogen atoms, and the angle formed by one hydrogen atom, oxygen, and the second hydrogen atom.
Subsequently, the bond distances and angles are rescaled to the interval $[-0.9, 0.9]$ for the $\cos^{-1}$ encoding (c.f. Figure~\ref{fig:pqc}), while the energies are scaled to the interval $[0, 1]$ for the output of the QNN. The dataset is divided into a training set consisting of 50 samples and a test set of 47 samples.
The QNN setup follows the procedure described in Section~\ref{sec:optimization}, employing 9 qubits and setting the ADAM learning rate to 0.01. The three input features are encoded cyclically, for instance, the first feature is encoded in the $\hat{R}_x$ gates of qubits 1, 4, and 7.

Figure~\ref{fig:h2o_results} depicts the loss functions $L_\text{var}$ and $L_\text{fit}$ as well as the number of shots utilized during the optimization of the QNN (\textbf{a}), along with the inference results for both training and test data (\textbf{b}). The optimization results closely resemble those shown in Figure~\ref{fig:opt_plot}, with the variance reduced by an order of magnitude and a substantial decrease in the fitting loss.
Moreover, the reduction in variance allows for a significant decrease in the number of shots required for gradient evaluation, while still yielding a lower fitting loss $L_\text{fit}$. The inference plots illustrate a significant reduction in the output variance through the variance regularization. Specifically, the width of the 95\% confidence interval decreases notably from an average value of 0.060 to 0.012 Hartree. Additionally, the averaged values of the R$^\text{2}$ scores demonstrate further improvement through the variance regularization.

\section{Results from the real hardware \label{sec:hardware}}

In this section we investigate the impact of the variance regularization on the performance of QNNs on real quantum computing hardware. 
All following computations are executed within the IBM Quantum ecosystem~\cite{IBMQuantum}.
Training a QNN on real quantum computing platforms remains a challenging and time-consuming task today.
In this section, we demonstrate that the reduced variance and the optimization procedure discussed in Section~\ref{sec:optimization} enable the optimization process on a real quantum computing backend, providing a notable performance boost due to the reduced number of shots.

\begin{figure}
\includegraphics[width=\linewidth]{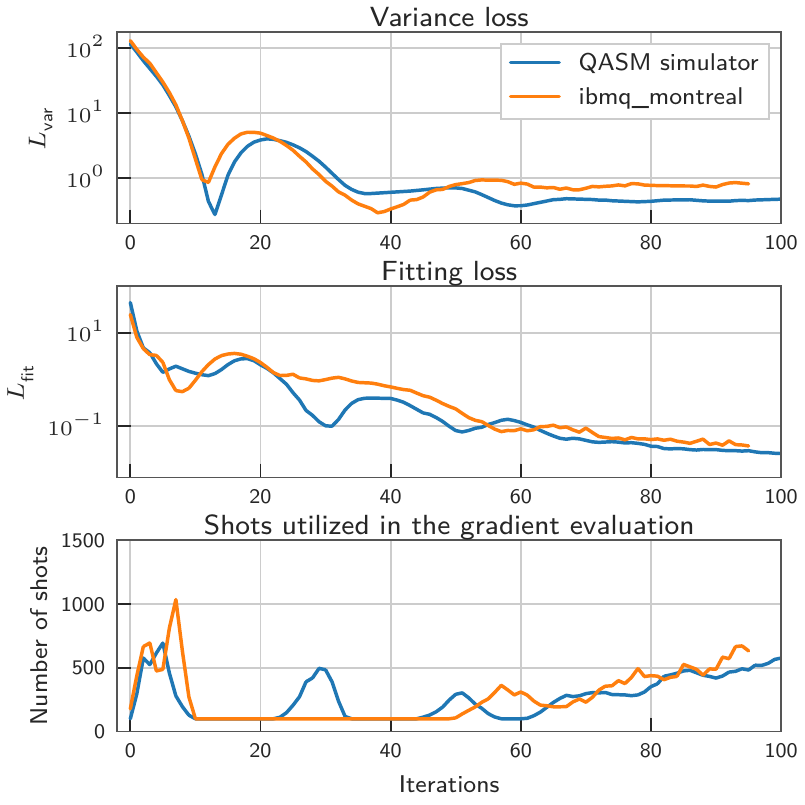}
\vspace{-0.6cm}
\caption{Results from the optimization on the IBM backend \texttt{ibmq\_montreal} and the corresponding shot-based QASM simulation, both utilizing 10 qubits. The optimization procedure is described in Section~\ref{sec:optimization}. \label{fig:opt_real_backend_100}}
\end{figure}

\begin{figure}
\includegraphics[width=\linewidth]{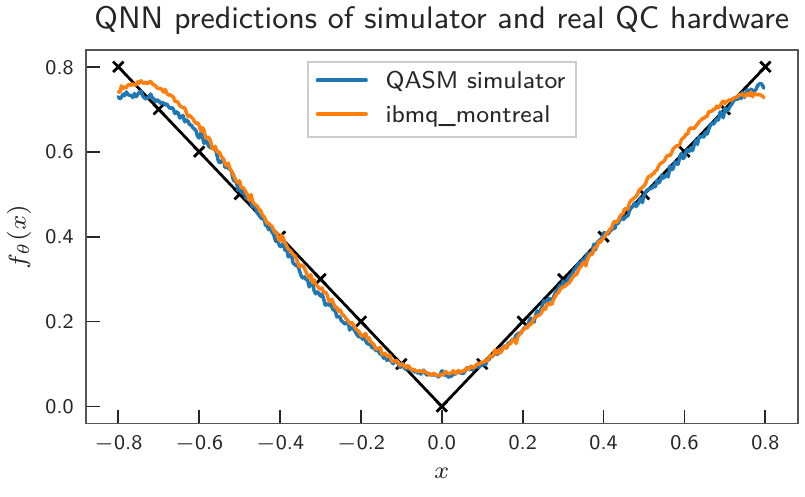}
\vspace{-0.6cm}
\caption{The output of the QNNs from both shot-based simulators (QASM) and real backends is presented.
The blue curve represents the result obtained from the converged QASM optimization.
The orange curve depicts the output after 95 training steps on the \texttt{ibmq\_montreal} backend.
It is worth noting that the results from the real quantum computing (QC) backend are obtained without the utilization of error mitigation techniques.
\label{fig:fig_real_backend_100}
} 
\end{figure}

We train the QNN using the procedure described in Section~\ref{sec:optimization}, wherein the last two-qubit gate $\hat{R}_{zz}$ of each layer is removed to achieve hardware-efficient linear entangling without introducing swapping gates.
To reduce computational time, we decrease the number of training points for fitting the data compared to the previous examples. The training data used in this example is shown in Figure~\ref{fig:fig_real_backend_100}.
Additionally, no error mitigation techniques are employed to expedite the process.
The qubit assignment remains fixed throughout the entire optimization process, as well as during inference.
However, evaluating the circuits resulting from the parameter-shift evaluation still consumes a significant amount of time.
It takes approximately 27 minutes to evaluate the 2618 circuits needed in this example for the parameter-shift derivative on the real backend, even with the lowest setting of 100 shots.
However, due to various reasons, such as pre- and post-processing of circuits and especially queuing, the training process for the QNN extends over a duration of several weeks.
Frequent re-calibration of the quantum hardware could potentially impact the optimization process, however, in our optimization this effect seems negligible.

Figure~\ref{fig:opt_real_backend_100} showcases the optimization performed on the IBM backend \texttt{ibmq\_montreal} for the absolute value function.
Additionally, we present the optimization with the exact same settings on the shot-based simulator.
Notably, both optimizations exhibit very similar behavior, indicating the potential for further reduction in the fitting loss on the real backend.
We anticipate that at some point, the additional noise introduced by the hardware will limit the further optimization process, resulting in a higher final loss compared to the shot-based simulation.
Due to the variance regularization, which greatly reduces the output variance of the QNN, we can perform the gradient evaluation with a relatively low number of shots.
This factor enables us to carry out the optimization of this example on the real backend in the first place. 
The full optimization curve of the shot-based QASM simulation is displayed in the Appendix in Section~\ref{sec:full_sim_opt}.

Figure~\ref{fig:fig_real_backend_100} displays the inference of the trained QNNs.
The blue curve is evaluated with the shot-based simulator and with parameters that are obtained from the converged QASM optimization displayed in Figure~\ref{fig:opt_real_backend_100}.
The orange curve is obtained from the IBM backend \texttt{ibmq\_montreal} with parameters that result from the optimization on the same backend.
At this stage of the optimziation, the output agrees well with the results from the shot-based simulator. 

We would like to emphasize that the optimization and inference on the real quantum computer were conducted without any error mitigation techniques.
Despite this, the QNN demonstrated a remarkable capability to adapt to hardware imperfections during real hardware training, as evidenced by its good agreement with the simulation results.
However, when evaluating the QNN on real backends using parameters obtained from the QASM optimization, we observed substantial deviations from the simulated outputs.
These findings strongly indicate that QNNs intended for evaluation on a quantum computer should be trained on the same machine.
Such an approach takes into account the specific hardware characteristics and imperfections, resulting in better performance and improved adaptability to the target quantum system.
In our view, this emphasizes the necessity for techniques that enable training directly on the hardware, instead of solely concentrating on simulations.

\begin{figure}
\vspace{2mm}
\includegraphics[width=\linewidth]{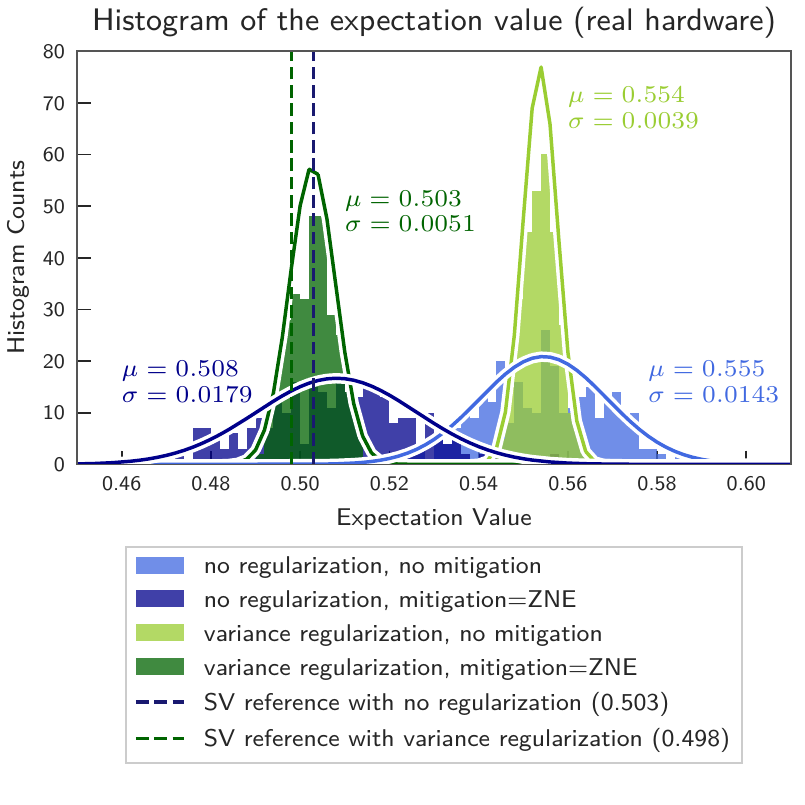}
\vspace{-0.6cm}
\caption{Histogram of the expectation value evaluated 300 times on the IBM backend \texttt{ibmq\_montreal}.
The darker colors are obtained with zero-noise extrapolation (ZNE).
Gaussian distributions are fitted to the obtained expectation values, the resulting mean ($\mu$) and the standard deviation ($\sigma$) are displayed in the same color. 
The dashed lines show the reference values obtained from the same QNNs evaluated on a noise-free statevector (SV) simulator. 
 \label{fig:real_histo} }
\end{figure}

The final example is motivated by Ref.~\cite{Wang.2024}, in which it is shown that the variance is increased by error mitigation.
Furthermore, it is discussed that error mitigation protocols can worsen the trainability of QNNs because of the increased variance. 
To investigate how the variance regularization is influence by the error mitigation, we compute results on the real backend utilizing the zero-noise extrapolation protocol~\cite{Temme.2017,Li.2017} as error mitigation. 
Zero-noise extrapolation involves deliberately amplifying the noise in the output by replicating gate operators in the circuit without altering its functionality.
This procedure allows the creation of a simple model based on the expectation value with controlled noise amplification, enabling extrapolation to the expectation value in the absence of any noise.

The QNNs in this example are obtained by an optimization with the shot-based simulator, both with and without variance regularization.
The training follows the protocol described above and in Section~\ref{sec:optimization} and fits the absolute value function.
The expectation values of the different QNNs are evaluated 300 times at $x=-0.5$ on the \texttt{ibmq\_montreal} backend utilizing 5000 shots in each run.
Figure~\ref{fig:real_histo} displays the histogram of the expectation values, both with and without zero-noise extrapolation.
Gaussian distributions are fitted to the 300 expectation values to illustrate the width of the distribution, and the mean ($\mu$) and the standard deviation ($\sigma$) of these Gaussians are provided.
It is evident that the variance of the expectation value is significantly improved by the variance regularization, regardless of the presence of zero-noise extrapolation.
Furthermore, the center of the expectation value is shifted to the noise-free reference value of the QNN by the application of zero-noise extrapolation.
The increased variance through the mitigation is visible for both QNNs, although it is not particularly significant in our example. Nonetheless, the variance reduction achieved through regularization remains strong even in the presence of zero-noise extrapolation.

\section{Conclusion}

In this work, we investigated the impact of finite sampling noise, an inevitable aspect of quantum computing, on QNNs. 
To mitigate this noise, we introduced a technique called variance regularization, which exploits the expressivity of QNNs to reduce the variance of the output. 
The method additionally includes the variance in the loss function that is minimized in the training. 
When the cost operator of the QNN is chosen to be diagonal in the computational basis (e.g. only using Pauli I and Z operators), the variance and its derivative can be computed from the same circuit evaluations as the function values and gradients.

We presented an example illustrating that noise-free QNNs obtained from simulators can exhibit a good fit but may suffer from a high variance, requiring either a huge number of shots or introducing significant amount of finite sampling noise.
We believe that this aspect is often underestimated in current research on QNNs.
Our findings demonstrate that the variance regularization significantly reduces the finite sampling noise.
Compared to results without regularization, the final QNN requires on average a magnitude fewer shots to achieve a similar level of noise.
We introduced an optimization procedure in which the contribution of the variance loss is adjusted during the optimization, resulting in substantial variance reduction and improved regression of the QNNs.
Empirically, we showed that the number of shots required during the gradient evaluation can be adjusted based on the variance of the fitting loss, leading to faster computation times.

In the final part of our study, we examined the impact of variance regularization on IBM's real quantum computing backends.
We demonstrated QNN optimization on a real backend, showcasing the adaptability of the QNN to hardware-specific characteristics during training.
The results show similar improvements in the variance reduction compared to the simulation examples.
Additionally, we observed that zero-noise extrapolation has no strong influence on the reduced variance of the output. 

We believe that variance regularization is a necessary step to make QNNs more practical for real-world applications, although the time-consuming training on real hardware remains a big challenge.
Exploring the application of variance regularization in other variational quantum algorithms may also be worth investigating when reducing finite sampling noise is crucial.

\begin{acknowledgments}
This work was supported by the German Federal Ministry of Education and Research through the project H2Giga-DegradEL3 (grant no. 03HY110D). 
We further acknowledge the use of IBM Quantum services for this work. The views expressed are those of the authors, and do not reflect the official policy or position of IBM or the IBM Quantum team.
\end{acknowledgments}

\appendix

\section{Appendix}

\subsection{Chebyshev with non-integer parameters \label{sub:cheb}}

\begin{figure}[h!]
\includegraphics[width=\linewidth]{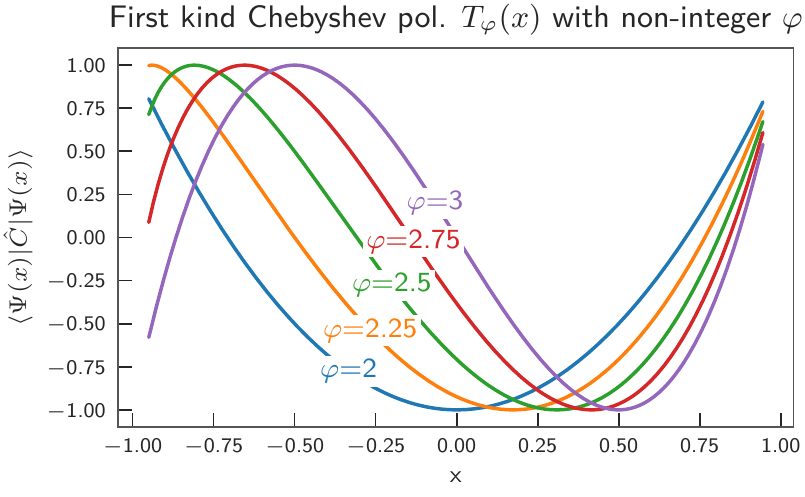}
\vspace{-0.6cm}
\caption{The first-kind Chebyshev polynomials $T_\varphi(x)$ with non-integer parameter. \label{fig:cheb_var}}
\end{figure}

The first-kind Chebyshev polynomials $T_\varphi(x)$ can be obtained from the following one qubit QNN: 
\begin{equation}
\ket{\Psi(x,\varphi)} = \hat{R}_x(\varphi \cos^{-1}(x)) 
\qquad \text{and} \qquad 
\hat{C} = \hat{Z}
\end{equation}
The cost operator is the Pauli Z operator. 
Figure~\ref{fig:cheb_var} shows the resulting curves for (non-integer) values of $\varphi \in [2,3]$. The original Chebyshev polynomials $T_2(x)$ and $T_3(x)$ are reproduced for $\varphi=2,3$.

\subsection{Dependencies on parameters in $\alpha(i)$ \label{sec:parameters}}

\begin{figure}[h!]
\vspace{2mm}
\includegraphics[width=1\linewidth]{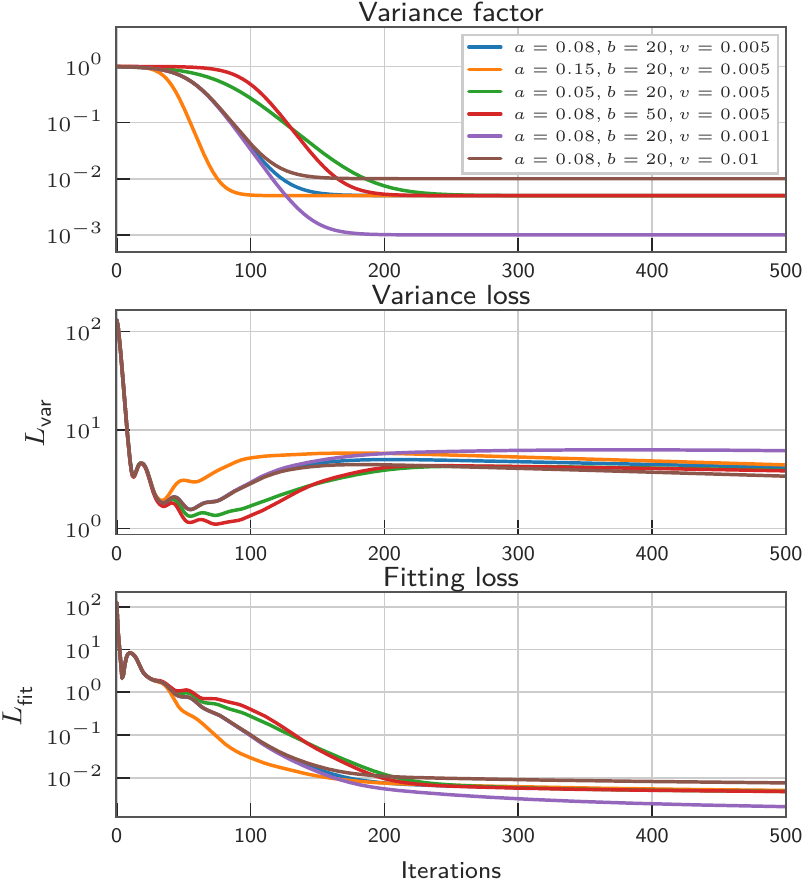}
\caption{Results for the noise-free optimization of the logarithm curve-fitting example detailed in Section~\ref{sec:optimization} with various hyper-parameters.
 \label{fig:diff_hyper_param}}
\end{figure}

In this section, we discuss the dependencies on the hyper-parameters in the function $\alpha_{a,b,v}(i)$ (cf. Eq. \eqref{eq:varfunc}) that is utilized in the variance regularization.
We execute the example of fitting the logarithmic function, as discussed in Section~\ref{sec:optimization}, with various hyper-parameters displayed in Figure~\ref{fig:varfac} on a noise-free simulator.
The function $\alpha_{a,b,v}(i)$, the variance loss $L_\text{var}$ and the fitting loss $L_\text{fit}$, are displayed in Figure~\ref{fig:diff_hyper_param}. We observe that although the progression of the variance loss and the fitting loss depends on the hyper-parameters, the final results consistently depend only on the value of the final plateau of $\alpha(i)$ set by parameter $v$. 
The hyper-parameter $v$ describes the proportion between the variance loss and the fitting loss; a large value reduces the variance more significantly at the cost of increasing the fitting loss.
In this work, we chose the same set of hyper-parameters for all discussed applications, but this might change for other applications. 
Here are some general intuitions on choosing the hyper-parameters: The plateau defined by the parameter $b$ should be sufficiently long to ensure that the variance is reduced if the fitting loss is significantly larger than $L_\text{var}$ at the beginning. The decay of $\alpha(i)$, controlled by the parameter $a$, influences the increase of the variance after reaching its minimum. However, the precise value of $a$ has minimal impact on the final result, as demonstrated in Figure~\ref{fig:diff_hyper_param}.

\subsection{Comparison between real quantum computing hardware and the fully converged simulated optimization \label{sec:full_sim_opt}}

\begin{figure}[h!]
\includegraphics[width=1\linewidth]{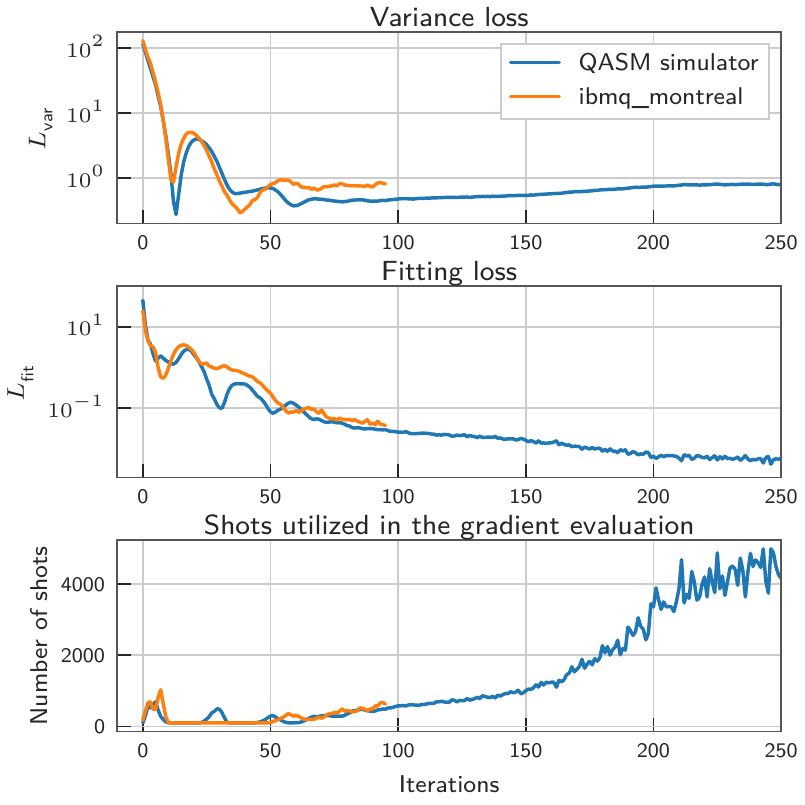}
\caption{Comparison between the optimization on the \texttt{ibmq\_montreal} backend and the fully converged simulated QASM optimziation. More details and discussion are given in Section~\ref{sec:hardware}.
 \label{fig:opt_real_backend_full}}
\end{figure}

Figure~\ref{fig:opt_real_backend_full} shows the comparison of the optimization on the real quantum computing (QC) backend \texttt{ibmq\_montreal} and the fully converged optimization carried out with the shot-based simulator. 
The optimization on the real backend was terminated at 95 iterations after running for several weeks. Although we were not able to run the optimization until convergence, the result shown in Figure~\ref{fig:opt_real_backend_full} (and Figure~\ref{fig:opt_real_backend_100}, respectively) is the longest cohesive optimization instance that we were able to run on real hardware, given the current practical limitations discussed in the main text.

\begin{thebibliography}{10}

\bibitem{Preskill.2018}
John Preskill.
\newblock ``Quantum computing in the nisq era and beyond''.
\newblock \href{https://dx.doi.org/10.22331/q-2018-08-06-79}{Quantum {\bf 2},
  79}~(2018).

\bibitem{Cerezo.2021}
M.~Cerezo, Andrew Arrasmith, Ryan Babbush, Simon~C. Benjamin, Suguru Endo,
  Keisuke Fujii, Jarrod~R. McClean, Kosuke Mitarai, Xiao Yuan, Lukasz Cincio,
  and Patrick~J. Coles.
\newblock ``Variational quantum algorithms''.
\newblock \href{https://dx.doi.org/10.1038/s42254-021-00348-9}{Nature Reviews
  Physics {\bf 3}, 625--644}~(2021).

\bibitem{Bharti.2022}
Kishor Bharti, Alba Cervera-Lierta, Thi~Ha Kyaw, Tobias Haug, Sumner
  Alperin-Lea, Abhinav Anand, Matthias Degroote, Hermanni Heimonen, Jakob~S.
  Kottmann, Tim Menke, Wai-Keong Mok, Sim Sukin, Leong-Chuan Kwek, and Al{\'a}n
  Aspuru-Guzik.
\newblock ``Noisy intermediate-scale quantum algorithms''.
\newblock \href{https://dx.doi.org/10.1103/RevModPhys.94.015004}{Rev. Mod.
  Phys. {\bf 94}, 015004}~(2022).

\bibitem{Scholkopf.2000}
Bernhard Sch\"{o}lkopf.
\newblock ``The kernel trick for distances''.
\newblock In Advances in Neural Information Processing Systems.
\newblock Volume~13, pages 301--307.
\newblock MIT Press~(2000).
\newblock
  url:~\url{https://proceedings.neurips.cc/paper_files/paper/2000/file/4e87337f366f72daa424dae11df0538c-Paper.pdf}.

\bibitem{Theodoridis.2008}
Sergios Theodoridis and Konstantinos Koutroumbas.
\newblock ``Pattern recognition''.
\newblock
  \href{https://dx.doi.org/10.1016/B978-1-59749-272-0.X0001-2}{{Academic
  Press}}. ~(2008).

\bibitem{Schuld.2019b}
Maria Schuld and Nathan Killoran.
\newblock ``Quantum machine learning in feature hilbert spaces''.
\newblock \href{https://dx.doi.org/10.1103/PhysRevLett.122.040504}{Phys. Rev.
  Lett. {\bf 122}, 040504}~(2019).

\bibitem{Lloyd.2020}
Seth Lloyd, Maria Schuld, Aroosa Ijaz, Josh Izaac, and Nathan Killoran.
\newblock ``Quantum embeddings for machine learning''~(2020).
\newblock  \href{https://doi.org/10.48550/arXiv.2001.03622}{arXiv:2001.03622}.

\bibitem{Havlicek.2019}
Vojt{\v{e}}ch Havl{\'i}{\v{c}}ek, Antonio~D. C{\'o}rcoles, Kristan Temme,
  Aram~W. Harrow, Abhinav Kandala, Jerry~M. Chow, and Jay~M. Gambetta.
\newblock ``Supervised learning with quantum-enhanced feature spaces''.
\newblock \href{https://dx.doi.org/10.1038/s41586-019-0980-2}{Nature {\bf 567},
  209--212}~(2019).

\bibitem{Hubregtsen.2022}
Thomas Hubregtsen, David Wierichs, Elies Gil-Fuster, Peter-Jan H.~S. Derks,
  Paul~K. Faehrmann, and Johannes~Jakob Meyer.
\newblock ``Training quantum embedding kernels on near-term quantum
  computers''.
\newblock \href{https://dx.doi.org/10.1103/PhysRevA.106.042431}{Phys. Rev. A
  {\bf 106}, 042431}~(2022).

\bibitem{Kubler.2021}
Jonas K\"{u}bler, Simon Buchholz, and Bernhard Sch\"{o}lkopf.
\newblock ``The inductive bias of quantum kernels''.
\newblock In Advances in Neural Information Processing Systems.
\newblock Volume~34, pages 12661--12673.
\newblock Curran Associates, Inc.~(2021).
\newblock
  url:~\url{https://proceedings.neurips.cc/paper_files/paper/2021/file/69adc1e107f7f7d035d7baf04342e1ca-Paper.pdf}.

\bibitem{Rapp.2024}
Frederic Rapp and Marco Roth.
\newblock ``Quantum gaussian process regression for bayesian optimization''.
\newblock \href{https://dx.doi.org/10.1007/s42484-023-00138-9}{Quantum Mach.
  Intell. {\bf 6}, 5}~(2024).

\bibitem{Mitarai.2018}
K.~Mitarai, M.~Negoro, M.~Kitagawa, and K.~Fujii.
\newblock ``Quantum circuit learning''.
\newblock \href{https://dx.doi.org/10.1103/PhysRevA.98.032309}{Phys. Rev. A
  {\bf 98}, 032309}~(2018).

\bibitem{Benedetti.2019}
Marcello Benedetti, Erika Lloyd, Stefan Sack, and Mattia Fiorentini.
\newblock ``Parameterized quantum circuits as machine learning models''.
\newblock \href{https://dx.doi.org/10.1088/2058-9565/ab4eb5}{Quantum Sci.
  Technol. {\bf 4}, 043001}~(2019).

\bibitem{Farhi.2018}
Edward Farhi and Hartmut Neven.
\newblock ``Classification with quantum neural networks on near term
  processors''~(2018).
\newblock  \href{https://doi.org/10.48550/arXiv.1802.06002}{arXiv:1802.06002}.

\bibitem{Cong.2019}
Iris Cong, Soonwon Choi, and Mikhail~D. Lukin.
\newblock ``Quantum convolutional neural networks''.
\newblock \href{https://dx.doi.org/10.1038/s41567-019-0648-8}{Nature Physics
  {\bf 15}, 1273--1278}~(2019).

\bibitem{Beer.2020}
Kerstin Beer, Dmytro Bondarenko, Terry Farrelly, Tobias~J. Osborne, Robert
  Salzmann, Daniel Scheiermann, and Ramona Wolf.
\newblock ``Training deep quantum neural networks''.
\newblock \href{https://dx.doi.org/10.1038/s41467-020-14454-2}{Nature
  communications {\bf 11}, 808}~(2020).

\bibitem{Zhang.2020}
Kaining Zhang, Min-Hsiu Hsieh, Liu Liu, and Dacheng Tao.
\newblock ``Toward trainability of quantum neural networks''~(2020).
\newblock  \href{https://doi.org/10.48550/arXiv.2011.06258}{arXiv:2011.06258}.

\bibitem{Cerezo.2022}
M.~Cerezo, Guillaume Verdon, Hsin-Yuan Huang, Lukasz Cincio, and Patrick~J.
  Coles.
\newblock ``Challenges and opportunities in quantum machine learning''.
\newblock \href{https://dx.doi.org/10.1038/s43588-022-00311-3}{Nature
  Computational Science {\bf 2}, 567--576}~(2022).

\bibitem{Wierichs.2020}
David Wierichs, Christian Gogolin, and Michael Kastoryano.
\newblock ``Avoiding local minima in variational quantum eigensolvers with the
  natural gradient optimizer''.
\newblock \href{https://dx.doi.org/10.1103/PhysRevResearch.2.043246}{Phys. Rev.
  Research {\bf 2}, 043246}~(2020).

\bibitem{Harrow.2021}
Aram~W. Harrow and John~C. Napp.
\newblock ``Low-depth gradient measurements can improve convergence in
  variational hybrid quantum-classical algorithms''.
\newblock \href{https://dx.doi.org/10.1103/PhysRevLett.126.140502}{Phys. Rev.
  Lett. {\bf 126}, 140502}~(2021).

\bibitem{Schuld.2019}
Maria Schuld, Ville Bergholm, Christian Gogolin, Josh Izaac, and Nathan
  Killoran.
\newblock ``Evaluating analytic gradients on quantum hardware''.
\newblock \href{https://dx.doi.org/10.1103/PhysRevA.99.032331}{Phys. Rev. A
  {\bf 99}, 032331}~(2019).

\bibitem{Wierichs.2022}
David Wierichs, Josh Izaac, Cody Wang, and Cedric Yen-Yu Lin.
\newblock ``General parameter-shift rules for quantum gradients''.
\newblock \href{https://dx.doi.org/10.22331/q-2022-03-30-677}{Quantum {\bf 6},
  677}~(2022).

\bibitem{Goto.2021}
Takahiro Goto, Quoc~Hoan Tran, and Kohei Nakajima.
\newblock ``Universal approximation property of quantum machine learning models
  in quantum-enhanced feature spaces''.
\newblock \href{https://dx.doi.org/10.1103/PhysRevLett.127.090506}{Phys. Rev.
  Lett. {\bf 127}, 090506}~(2021).

\bibitem{Schuld.2021}
Maria Schuld, Ryan Sweke, and Johannes~Jakob Meyer.
\newblock ``Effect of data encoding on the expressive power of variational
  quantum-machine-learning models''.
\newblock \href{https://dx.doi.org/10.1103/PhysRevA.103.032430}{Phys. Rev. A
  {\bf 103}, 032430}~(2021).

\bibitem{PerezSalinas.2021}
Adri{\'a}n P{\'e}rez-Salinas, David L{\'o}pez-N{\'u}{\~n}ez, Artur
  Garc{\'i}a-S{\'a}ez, P.~Forn-D{\'i}az, and Jos{\'e}~I. Latorre.
\newblock ``One qubit as a universal approximant''.
\newblock \href{https://dx.doi.org/10.1103/PhysRevA.104.012405}{Phys. Rev. A
  {\bf 104}, 012405}~(2021).

\bibitem{Vidal.2020}
Francisco Javier~Gil Vidal and Dirk~Oliver Theis.
\newblock ``Input redundancy for parameterized quantum circuits''.
\newblock \href{https://dx.doi.org/10.3389/fphy.2020.00297}{Front. Phys. {\bf
  8}, 297}~(2020).

\bibitem{PerezSalinas.2020}
Adri{\'a}n P{\'e}rez-Salinas, Alba Cervera-Lierta, Elies Gil-Fuster, and
  Jos{\'e}~I. Latorre.
\newblock ``Data re-uploading for a universal quantum classifier''.
\newblock \href{https://dx.doi.org/10.22331/q-2020-02-06-226}{Quantum {\bf 4},
  226}~(2020).

\bibitem{Caro.2022}
Matthias~C. Caro, Hsin-Yuan Huang, M.~Cerezo, Sharma Kunal, Andrew Sornborger,
  Lukasz Cincio, and Patrick~J. Coles.
\newblock ``Generalization in quantum machine learning from few training
  data''.
\newblock \href{https://dx.doi.org/10.1038/s41467-022-32550-3}{Nature
  communications {\bf 13}, 4919}~(2022).

\bibitem{McClean.2018}
Jarrod~R. McClean, Sergio Boixo, Vadim~N. Smelyanskiy, Ryan Babbush, and
  Hartmut Neven.
\newblock ``Barren plateaus in quantum neural network training landscapes''.
\newblock \href{https://dx.doi.org/10.1038/s41467-018-07090-4}{Nature
  communications {\bf 9}, 4812}~(2018).

\bibitem{Arrasmith.2021}
Andrew Arrasmith, M.~Cerezo, Piotr Czarnik, Lukasz Cincio, and Patrick~J.
  Coles.
\newblock ``Effect of barren plateaus on gradient-free optimization''.
\newblock \href{https://dx.doi.org/10.22331/q-2021-10-05-558}{Quantum {\bf 5},
  558}~(2021).

\bibitem{Holmes.2022}
Zo{\"e} Holmes, Sharma Kunal, M.~Cerezo, and Patrick~J. Coles.
\newblock ``Connecting ansatz expressibility to gradient magnitudes and barren
  plateaus''.
\newblock \href{https://dx.doi.org/10.1103/PRXQuantum.3.010313}{PRX Quantum
  {\bf 3}, 010313}~(2022).

\bibitem{Marrero.2021}
Carlos~Ortiz Marrero, M{\'a}ria Kieferov{\'a}, and Nathan Wiebe.
\newblock ``Entanglement-induced barren plateaus''.
\newblock \href{https://dx.doi.org/10.1103/PRXQuantum.2.040316}{PRX Quantum
  {\bf 2}, 040316}~(2021).

\bibitem{Cerezo.2021b}
M.~Cerezo, Akira Sone, Tyler Volkoff, Lukasz Cincio, and Patrick~J. Coles.
\newblock ``Cost function dependent barren plateaus in shallow parametrized
  quantum circuits''.
\newblock \href{https://dx.doi.org/10.1038/s41467-021-21728-w}{Nature
  communications {\bf 12}, 1791}~(2021).

\bibitem{Uvarov.2021}
A.~V. Uvarov and J.~D. Biamonte.
\newblock ``On barren plateaus and cost function locality in variational
  quantum algorithms''.
\newblock \href{https://dx.doi.org/10.1088/1751-8121/abfac7}{J. Phys. A: Math.
  Theor. {\bf 54}, 245301}~(2021).

\bibitem{Pesah.2021}
Arthur Pesah, M.~Cerezo, Samson Wang, Tyler Volkoff, Andrew~T. Sornborger, and
  Patrick~J. Coles.
\newblock ``Absence of barren plateaus in quantum convolutional neural
  networks''.
\newblock \href{https://dx.doi.org/10.1103/PhysRevX.11.041011}{Phys. Rev. X
  {\bf 11}, 041011}~(2021).

\bibitem{Larocca.2023}
Mart{\'i}n Larocca, Nathan Ju, Diego Garc{\'i}a-Mart{\'i}n, Patrick~J. Coles,
  and Marco Cerezo.
\newblock ``Theory of overparametrization in quantum neural networks''.
\newblock \href{https://dx.doi.org/10.1038/s43588-023-00467-6}{Nature
  Computational Science {\bf 3}, 542--551}~(2023).

\bibitem{Wang.2021}
Samson Wang, Enrico Fontana, M.~Cerezo, Kunal Sharma, Akira Sone, Lukasz
  Cincio, and Patrick~J. Coles.
\newblock ``Noise-induced barren plateaus in variational quantum algorithms''.
\newblock \href{https://dx.doi.org/10.1038/s41467-021-27045-6}{Nature
  communications {\bf 12}, 6961}~(2021).

\bibitem{Wang.2024}
Samson Wang, Piotr Czarnik, Andrew Arrasmith, M.~Cerezo, Lukasz Cincio, and
  Patrick~J. Coles.
\newblock ``Can error mitigation improve trainability of noisy variational
  quantum algorithms?''.
\newblock \href{https://dx.doi.org/10.22331/q-2024-03-14-1287}{Quantum {\bf 8},
  1287}~(2024).

\bibitem{braket_pricing}
{Amazon Web Services}.
\newblock ``Quantum computer and simulator – amazon braket pricing – aws''.
\newblock \url{https://aws.amazon.com/de/braket/pricing/}.
\newblock Accessed: 2023-04-11.

\bibitem{Skolik.2021}
Andrea Skolik, Jarrod~R. McClean, Masoud Mohseni, Patrick {van der Smagt}, and
  Martin Leib.
\newblock ``Layerwise learning for quantum neural networks''.
\newblock \href{https://dx.doi.org/10.1007/s42484-020-00036-4}{Quantum Mach.
  Intell. {\bf 3}, 5}~(2021).

\bibitem{Kyriienko.2021}
Oleksandr Kyriienko, Annie~E. Paine, and Vincent~E. Elfving.
\newblock ``Solving nonlinear differential equations with differentiable
  quantum circuits''.
\newblock \href{https://dx.doi.org/10.1103/PhysRevA.103.052416}{Phys. Rev. A
  {\bf 103}, 052416}~(2021).

\bibitem{Somma.2002}
R.~Somma, G.~Ortiz, J.~E. Gubernatis, E.~Knill, and R.~Laflamme.
\newblock ``Simulating physical phenomena by quantum networks''.
\newblock \href{https://dx.doi.org/10.1103/PhysRevA.65.042323}{Phys. Rev. A
  {\bf 65}, 042323}~(2002).

\bibitem{Guerreschi.2017}
Gian~Giacomo Guerreschi and Mikhail Smelyanskiy.
\newblock ``Practical optimization for hybrid quantum-classical
  algorithms''~(2017).
\newblock  \href{https://doi.org/10.48550/arXiv.1701.01450}{arXiv:1701.01450}.

\bibitem{Abramowitz.1988}
Milton Abramowitz, Irene~A. Stegun, and Robert~H. Romer.
\newblock ``Handbook of mathematical functions with formulas, graphs, and
  mathematical tables''.
\newblock \href{https://dx.doi.org/10.1119/1.15378}{American Journal of Physics
  {\bf 56}, 958}~(1988).

\bibitem{Treinish.2023}
Ali Javadi-Abhari, Matthew Treinish, Kevin Krsulich, Christopher~J. Wood, Jake
  Lishman, Julien Gacon, Simon Martiel, Paul~D. Nation, Lev~S. Bishop,
  Andrew~W. Cross, Blake~R. Johnson, and Jay~M. Gambetta.
\newblock ``Quantum computing with qiskit''~(2024).
\newblock  \href{https://doi.org/10.48550/arXiv.2405.08810}{arXiv:2405.08810}.

\bibitem{Kreplin.2023}
David~A. Kreplin, Moritz Willmann, Jan Schnabel, Frederic Rapp, Manuel
  Hagelüken, and Marco Roth.
\newblock ``squlearn -- a python library for quantum machine learning''~(2024).
\newblock  \href{https://doi.org/10.48550/arXiv.2311.08990}{arXiv:2311.08990}.

\bibitem{Benaroya.2005}
Haym Benaroya, Seon~Mi Han, and Mark Nagurka.
\newblock ``Probability models in engineering and science''.
\newblock \href{https://dx.doi.org/10.1201/9781420056341}{{CRC press}}.
  ~(2005).

\bibitem{Kingma.2017}
Diederik~P. Kingma and Jimmy Ba.
\newblock ``Adam: A method for stochastic optimization''~(2017).
\newblock  \href{https://doi.org/10.48550/arXiv.1412.6980}{arXiv:1412.6980}.

\bibitem{Kiss.2022}
Oriel Kiss, Francesco Tacchino, Sofia Vallecorsa, and Ivano Tavernelli.
\newblock ``Quantum neural networks force fields generation''.
\newblock \href{https://dx.doi.org/10.1088/2632-2153/ac7d3c}{Mach. Learn.: Sci.
  Technol. {\bf 3}, 035004}~(2022).

\bibitem{water_dataset}
Gunnar~Alexander Schmitz.
\newblock ``water-dz-f12-static-g32n-1m\_pes-dzero.xyz''.
\newblock In Machine Learning for Potential Energy Surface Construction: A
  Benchmark Set.
\newblock \href{https://dx.doi.org/10.7910/DVN/C9ISSX/T3ROI5}{Harvard
  Dataverse}~(2019).

\bibitem{Schmitz.2019}
Gunnar Schmitz, Ian~Heide Godtliebsen, and Ove Christiansen.
\newblock ``Machine learning for potential energy surfaces: An extensive
  database and assessment of methods''.
\newblock \href{https://dx.doi.org/10.1063/1.5100141}{J. Chem. Phys. {\bf 150},
  244113}~(2019).

\bibitem{IBMQuantum}
{IBM Quantum}.
\newblock \url{https://quantum-computing.ibm.com/}~(2023).

\bibitem{Temme.2017}
Kristan Temme, Sergey Bravyi, and Jay~M. Gambetta.
\newblock ``Error mitigation for short-depth quantum circuits''.
\newblock \href{https://dx.doi.org/10.1103/PhysRevLett.119.180509}{Phys. Rev.
  Lett. {\bf 119}, 180509}~(2017).

\bibitem{Li.2017}
Ying Li and Simon~C. Benjamin.
\newblock ``Efficient variational quantum simulator incorporating active error
  minimization''.
\newblock \href{https://dx.doi.org/10.1103/PhysRevX.7.021050}{Phys. Rev. X {\bf
  7}, 021050}~(2017).

\end{thebibliography}

\end{document}